# A coherent way to image dislocations

V.L.R. Jacques[1,2,3], S. Ravy[2], D. Le Bolloc'h[1], E. Pinsolle[1,2], M. Sauvage-Simkin[2] & F. Livet[4]

[1]*Laboratoire de Physique des Solides, CNRS, UMR8502, Université Paris-Sud, 91405 Orsay Cedex, France*
[2]*Synchrotron SOLEIL, L'Orme des Merisiers, Saint-Aubin, BP43, 91192 Gif-sur-Yvette Cedex, France*
[3]*European Synchrotron Radiation Facility, 6 rue Jules Horowitz, BP220, 38043 Grenoble Cedex, France*
[4]*Sciences et Ingénierie des Matériaux et Procédés, INP Grenoble CNRS UJF, BP 75, 38402 St Martin d'Hères Cedex, France*

**The use of coherent x-ray beams has been greatly developing for the past decades[1]. They are now used by a wide scientific community to study biological materials[2], phase transitions in crystalline materials[3], soft matter[4], magnetism[5-6], strained structures[7-8], or nano-objects[9]. Different kinds of measurements can be carried out: x-ray photon correlation spectroscopy[10] allowing studying dynamics in soft[11] and hard[12] matter, and coherent diffraction imaging[13] enabling to reconstruct the shape[14] and strain[15] of some objects by using methods such as holography[16] or ptychography[17]. In this article, we show that coherent x-ray diffraction[18] (CXRD) brings a new insight in another scientific field: the detection of single phase defects in bulk materials. Extended phase objects such as dislocations embedded in the bulk are usually probed by electron microscopy or X-ray topography. However, electron microscopy is restricted to thin samples, and x-ray topography is resolution-limited. We show here that CXRD brings much more accurate information about dislocation lines (DLs) in bulk samples and opens a route for a better understanding of the fine structure of the core of bulk dislocations.**

Understanding the origin and behaviour of topological defects in hard condensed matter is a challenging problem. Their study has always been of prime importance because of the large influence they have on the physical properties of materials[19]. Good examples are the plasticity of crystalline materials, determined by the mobility of dislocations, the electronic conductivity, strongly affected by the density of dislocations, or even more complex phenomena like the sliding motion of charge/spin density waves (CDW/SDW) mainly related to the dynamics of climbing DW dislocations. Only a few techniques, such as x-ray topography[20] or electron microscopy[21] are able to probe the properties and dynamics of dislocations. We show that CXRD not only is a new tool to image DLs located deep inside samples, but also provides complementary information to the so far existing techniques. Another strength of CXRD is its ability to probe phase defects of any x-ray diffractive order parameter. Its sensitivity to CDW[22] and SDW[6] dislocations has already been reported. In addition, combination of CXRD with topography proves efficient to get rich and accurate information of extended bulk dislocations and allows to directly image them.

The combination of CXRD and topography takes advantage of the behaviour of a light beam limited in size by slits. If an aperture *a* defines the size of a λ-wavelength monochromatic beam, the latter keeps parallel up to a distance $d_{FF}=a^2/4\lambda$, and then diverges linearly. If both sample and detector are inserted in the parallel beam regime, *i.e.* below a distance $d_{FF}$ from the slits, a topography image is obtained. In practice, this corresponds to large apertures *a*. On the contrary, a CXRD image is obtained when the sample receives a parallel beam while the detector receives a diverging beam *i.e.* if $d_{FF}$ is intermediate between sample and detector distances from the slits, and when the aperture *a* is lower than the transverse coherence length $\xi_T$ of the beam. The two setups are displayed in Fig.1 (see section Methods for further details about the setups).

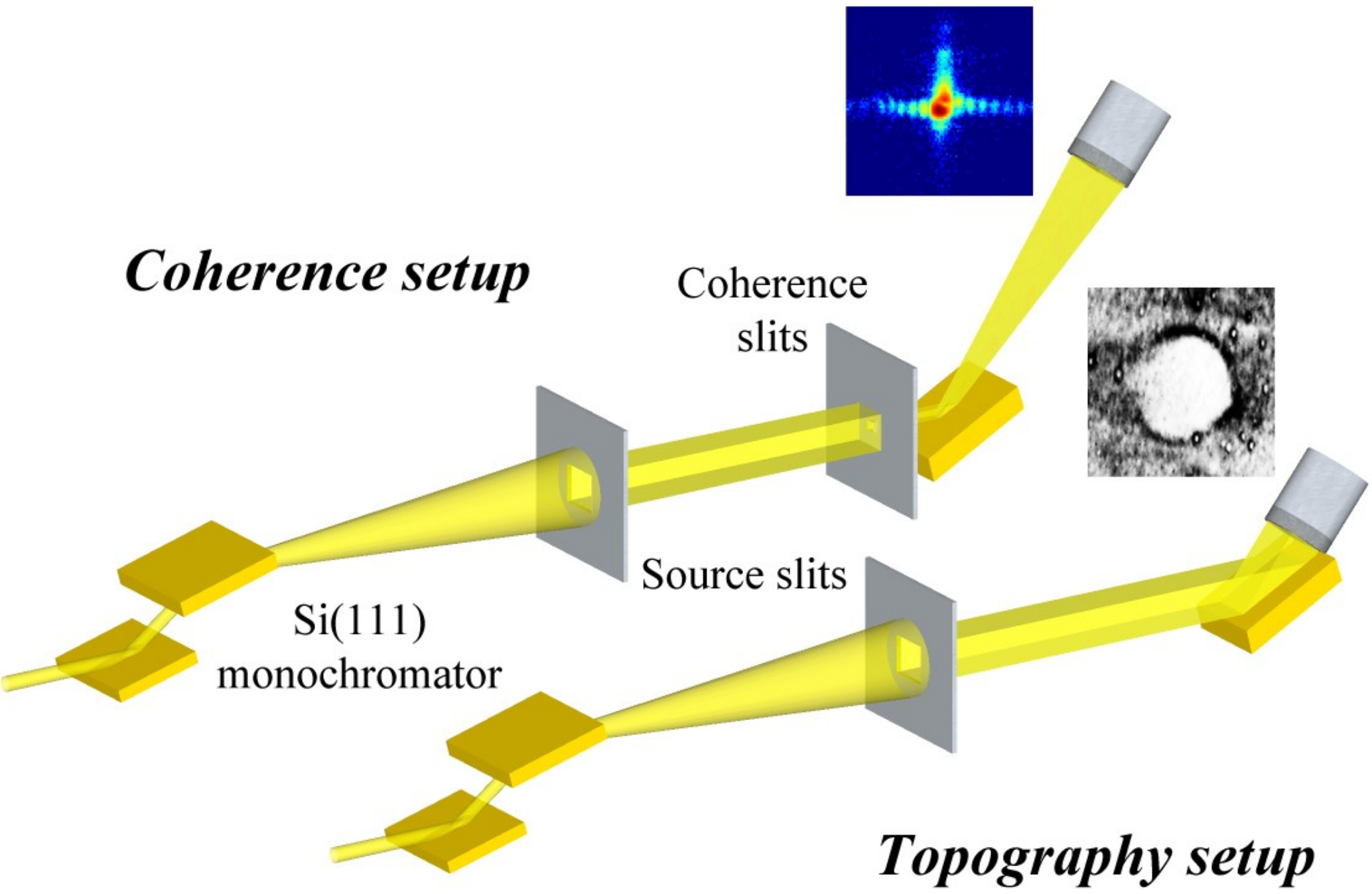

**Figure 1 | Coherence and topography setups used for the experiment**. A monochromatic beam was selected by a double-crystal Si(111) monochromator. No focusing has been used to get the required parallel beam for topography. Source slits were used as secondary source. For topography, attenuators were also introduced. For coherence, slits were used before the sample to select only the coherent part of the beam. See section Methods for further information.

To demonstrate the strength of combining CXRD and topography, we present a study of the well-known oxygen-induced dislocations in silicon[20]. This sample processing stabilizes two kinds of defects: stacking faults (SF) bounded by partial dislocation loops (Burgers vectors $\mathbf{b_S}$=1/3<111>), and perfect dislocation loops, with Burgers vectors $\mathbf{b_P}$=1/2<110>. As shown in Fig. 2a, the latter are dissociated into two partial dislocations, with Burgers vectors $\mathbf{b_1}$ and $\mathbf{b_2}$ respectively, so that $\mathbf{b_P} = \mathbf{b_1} + \mathbf{b_2}$. The partial dislocations are called glissile if their direction and Burgers vector belong to the same {111} glide plane, as sketched in Fig. 2a, or sessile, if not. As a consequence, a SF stands between the two partial DLs[23] provided one at least is glissile. Topography is first used to locate the defects to be studied. As the dislocation loops are extended over 100 µm on the average, the several hundred micrometers size beam used in topography is very convenient to get a rapid overview of the defects positions and concentration. A 5 µm beam is then used to make a CXRD study of the selected defect.

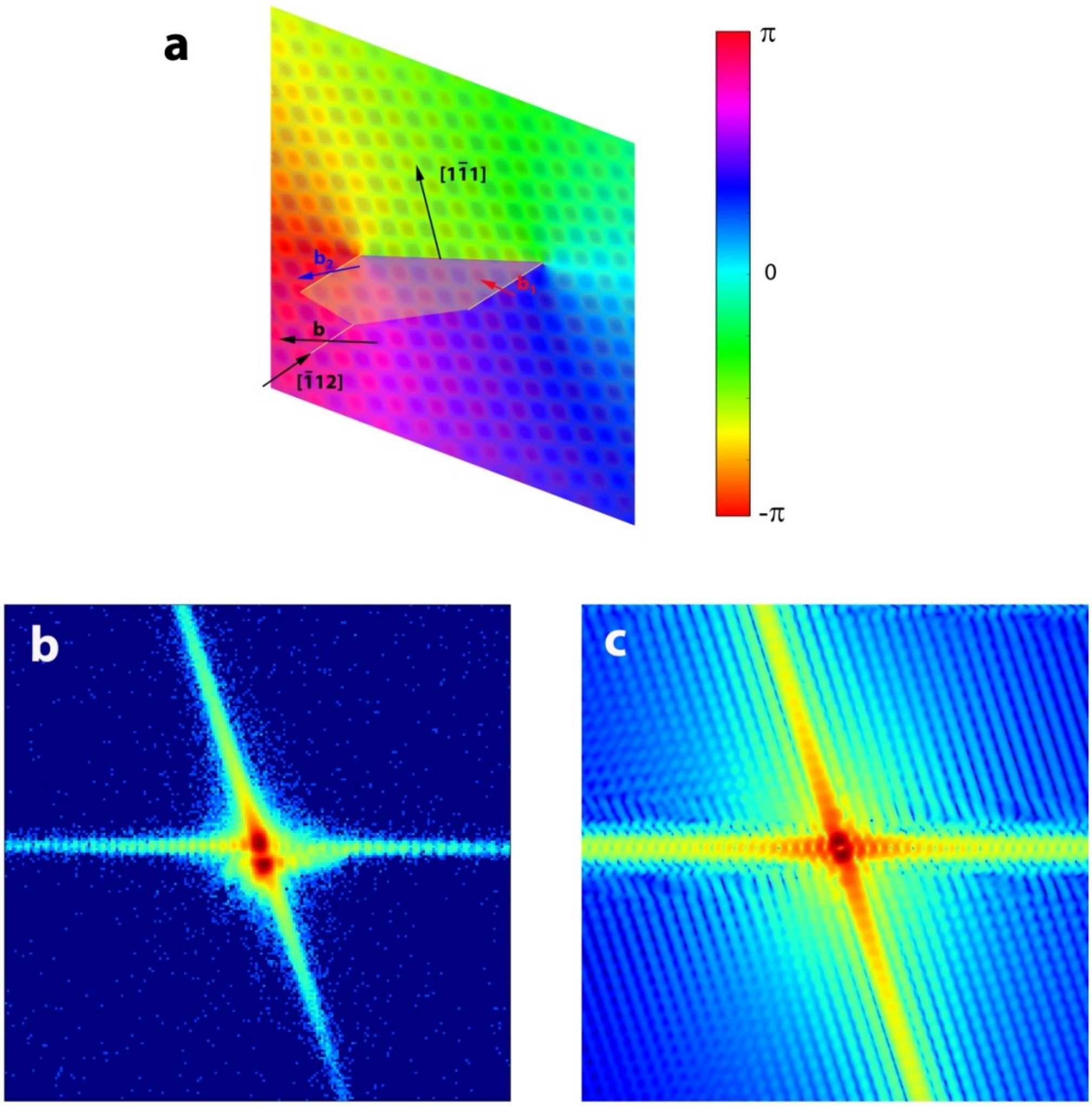

**Figure 2 | Coherent Bragg reflection associated to a dissociated dislocation**. **a**, Dissociation of a perfect **b**=1/2[110] dislocation developing along a [-112] direction into two glissile partials: $\mathbf{b_1}$=1/6[121] and $\mathbf{b_2}$=1/6[21-1]. The yellow plane is a stacking fault. Superimposed to the schematic atomic lattice, the colour scale refers to the phase shift induced by the presence of the dissociated dislocation; **b,** measured (220) Bragg reflection in the presence of a dissociated dislocation. A splitting due to the phase shift is observed along with a streak perpendicular to the stacking fault; **c,** Calculated diffraction pattern associated to a dissociated dislocation, in good agreement with the measured intensity distribution shown in **b**.

Thanks to its high sensitivity to phase shifts compared to standard diffraction, CXRD is a very powerful technique to probe dislocations. The π phase shift induced by a dislocation is expected to lead to a destructive interference at the Bragg position. In the case of a dislocation located in the middle of the coherently illuminated volume, the Bragg peak is expected to split into two parts, each of them having the same angular width as that of a Bragg reflection associated to a perfect lattice, and approximately half its maximum intensity (for more details, see Supplementary Discussion 1 in Supplementary Information). This dislocation induced splitting has been experimentally recorded on the (220) Bragg reflection, as shown in Fig. 2b. In addition, despite the thinness of the nm-size SFs due to dissociation, the data reveal streaks appearing together with the splitting, along <111> directions, *i.e.* perpendicular to the SF. A calculation of the expected (220) reflection in presence of a dissociated dislocation shows a very good agreement with the measured intensity distribution (Fig. 2c).

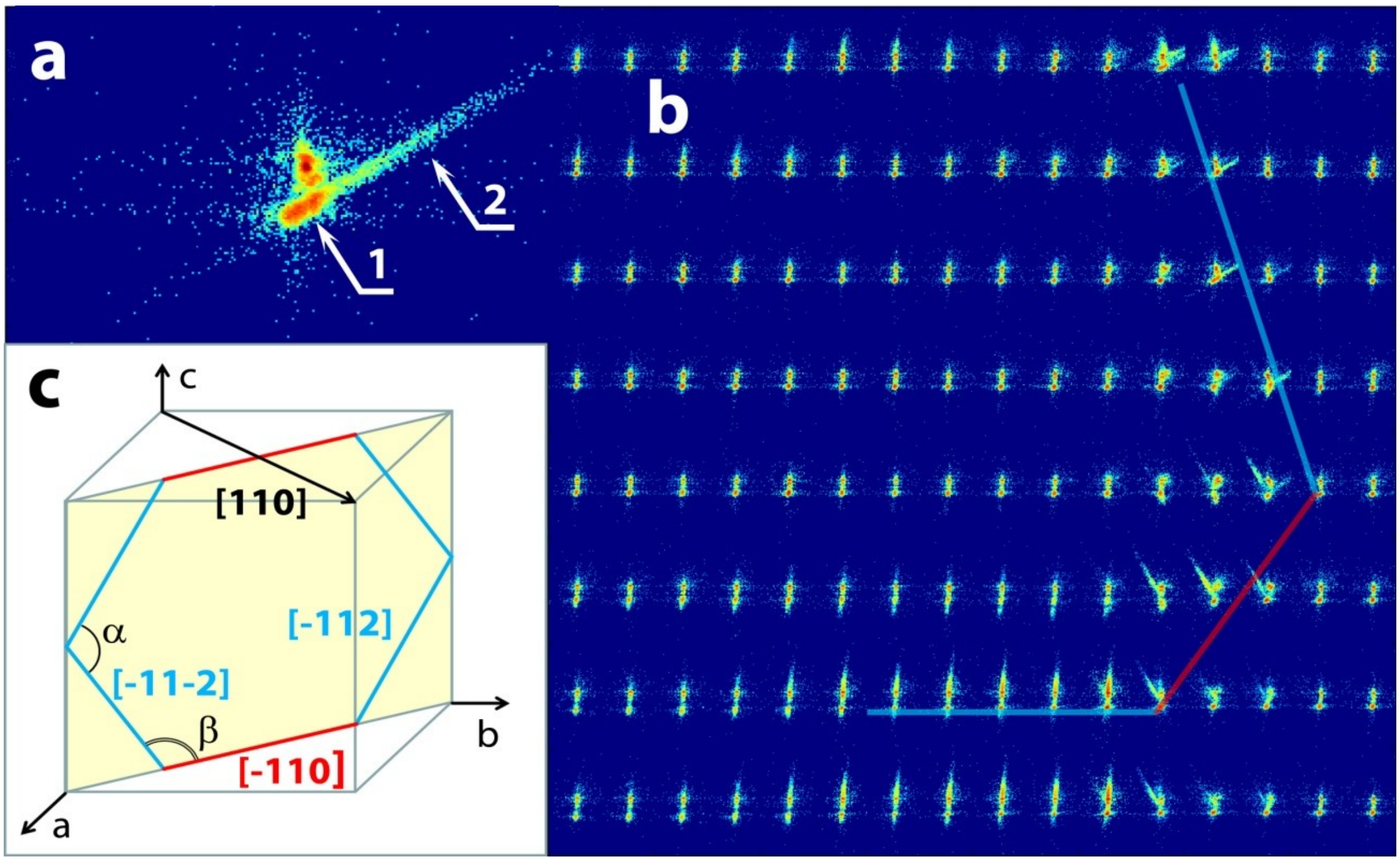

**Figure 3 | Principle of the scanning CXRD. a**, Typical coherent diffraction pattern obtained on the (220) reflection in presence of a dissociated DL, with a 10 mdeg detuning of the sample. The truncation rod is split (1) due to the phase shift induced by the dislocation, and a diffuse streak (2) appears along the splitting, perpendicular to the SF located in between the dissociated lines. **b**, The typical patterns shown in (a) only appear in the close vicinity of the DL. The diffuse streak is always perpendicular to the DL (and SF) direction. The red lines correspond to DLs along the [-110] direction (sessile), and the blue ones along [-11±2] directions (glissile). **c**, The measured DLs directions are in agreement with the expected directions and angles for dissociated dislocations ($\alpha = 109.47^o$, $\beta = 125.26^o$).

This unique signature under a coherent light allows imaging such extended defects by performing a spatial scan across the area of the dislocation, and taking advantage of the difference in intensity at the Bragg position when the DL is or is not in the illuminated volume. When a perfect region is probed, the reflection is unique, and a maximum appears, whereas when the DL is in the illuminated volume, the peak is split, and the value at the Bragg position is lowered. An enhanced contrast is obtained by slightly detuning the sample to avoid the intense Bragg maximum. The measured distribution of intensity is then composed of a cut through the truncation rods (TRs) associated to the main (220) Bragg reflection, and to the diffracting slit fringes (for more details, see Supplementary Discussion 2 in Supplementary Information). In that situation, Bragg reflection splitting is observed on the proper associated TR (label 1 in Fig. 3a) and the signal from the diffuse streak becomes thereby more contrasted (label 2 in Fig. 3a). Fig. 3b displays a typical 2D scan performed with a 5 μm coherent beam over a region beforehand selected with topography. As expected, the splitting and streaks only appear in the close vicinity of the DLs, which is the key to perform imaging of the line with beam-size resolution. The DLs directions are determined by delimiting in space the regions where the splitting and streaks occur, and are found to be the expected ones for a perfect dislocation loop with $\mathbf{b_p}$=1/2[110] (see Fig. 3c). Moreover, the streaks appear systematically perpendicular to the partial dislocations possible {111} glide planes. The same method was repeated across another dislocation loop (see Fig. 4a), first selected with topography (Fig. 4b). A 200 x 200 $\mu m^2$ region was scanned making steps having the same size as the beam footprint in each direction. The blue valley of Bragg peak minima in Fig. 4c. directly gives the position of the dislocation loop lines and is in good agreement with the topography image. However, the present technique allows going beyond topography, as it informs about the presence of a dissociation of the DL if any, through the local measurement of additional streaks in the diffraction pattern.

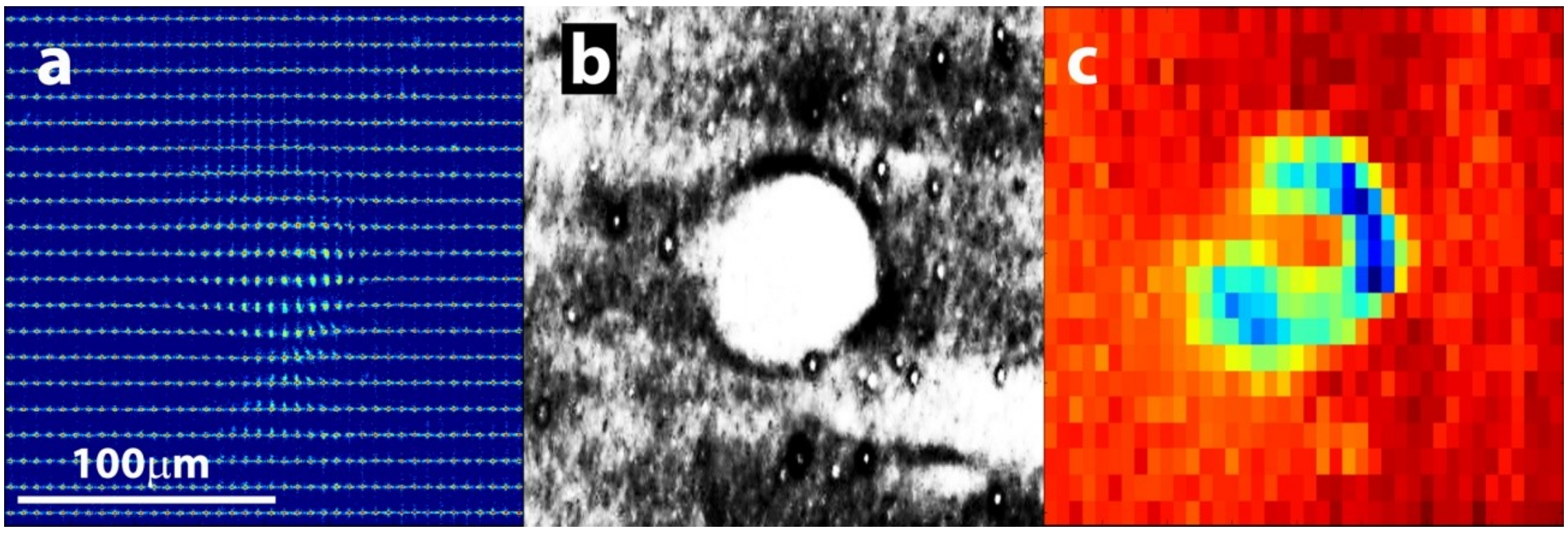

**Figure 4 | Image of a dislocation loop obtained with scanning CXRD. a**, Coherent diffraction patterns obtained in the area of a dislocation loop, beforehand selected making **b**, a topography image of the same dislocation loop. **c**, Imaging performed taking the intensity at Bragg reflection position for each pattern shown in **a**. When the DL is in the probed volume, the value is minimum, which allows following it in space. The perfectly ordered regions appear in red, the DL in blue. The topography and CXRD images are in good agreement.

In conclusion, this article proves the high sensitivity of CXRD to any dislocation embedded within a few micrometers in the bulk, and shows that direct imaging of bulk dislocations by CXRD is possible. The combination of CXRD and topography is a very flexible method which can be used at every beamline dedicated to diffraction, to first locate the object in the real space and second to probe its fine structure through its Fourier transform in the reciprocal space.

In the canonical example of dislocation loops in silicon, the obtained CXRD patterns display a destructive interference *at* the Bragg position as well as a diffuse streak emanating from the fine structure of the dislocation core region. The diffraction pattern is compatible with the presence of dissociated dislocations, as observed by electron microscopy studies in thin samples. Only the destructive interference allows performing the imaging of the DL itself, which highlights the need of a coherent beam for this study.

To push the study further, such improvements as 3D data collection, the use of sub-micrometer beams[24,25] and of reconstruction algorithms *e.g.* ptychography would give access to unprecedented resolution and to the strain field associated to the dislocation line. Besides, it also represents a unique opportunity to probe the core of topological objects embedded in the bulk, with an unequalled resolution and in a non destructive way. In that case, a more detailed model would have to be used to describe the dislocation core with more accuracy, although in the present case the simple phase model nicely explains the different features of the diffraction pattern.

With the quest for new materials like SiC and the advent of new processes like bonding in view of improving device performances, the study of topological defects, in particular in semiconductors, is experiencing a revival for which this work opens significant perspectives. This approach also benefits from the easiness to use various sample environments, such as high pressure[26], AFM[27], magnetic/electric fields[28] or Laser pulses[29]. For these in-situ experiments, the new XFEL sources may enable to probe fast dynamics of dislocations at the femtosecond timescale.

## METHODS

### Experimental details

Preliminary experiments were performed at the BM02 beamline of the ESRF. The data presented here were obtained during an experiment performed at the CRISTAL beamline of the synchrotron-SOLEIL. The source is a U20 undulator, and a Si(111) double-crystal monochromator (DCM) selected a wavelength $\lambda$=1.7462 Å, with $\Delta\lambda/\lambda$=1.4 $10^{-4}$. Focusing devices (DCM sagittal bender and focusing mirrors) were not used in order to have a parallel beam for topography. Rejection of harmonics was ensured by detuning the DCM at about 50% of the maximum value of the rocking curve. Secondary slits were used for beam collimation 13 m upstream of the sample and attenuators prevented saturation of the detector. The coherent beam was generated by setting the secondary slits to 200 µm, leading to a transverse coherence length of 11.3 µm at the sample position. Detection was performed with a 13 µm pixel size ANDOR direct illumination CCD camera standing 2.2 m after the sample. Slits located 8 cm upstream of the sample were closed at 7 µm, allowing to get a parallel beam at the sample position, and a divergent one at the detector position. The total transverse coherence degree[18] was then about 80%.

**Sample preparation**

The 18x9x0.5 $mm^3$ silicon single crystal was grown by the Czochralski method, and then annealed during 35 hours at 1100 °C under an oxygen atmosphere. The concentration in oxygen impurities was then $10^{18}$ atoms/$cm^3$. The surface was oriented (110). A 4 mdeg mosaicity was measured by performing rocking scans on the (220) reflection. The sample was provided by G. Rolland from CEA-LETI Grenoble.

# Supplementary Discussion 1
# Destructive interference due to the presence of a dislocation line in a coherently illuminated volume

The problem can be first simplified to a one dimensional model. If one considers a perfectly ordered lattice (blue line on Supplementary Figure 1.a), the associated phase is constant in space, and can be arbitrarily chosen to be 0 (blue line on Supplementary Figure 1.b). Any Bragg reflection associated to this lattice is unique, and its width is inversely proportional to the dimension of the system (blue curve on Supplementary Figure 1.c). By introducing a π phase shift in this lattice, one obtains the lattice and its phase variations shown with red lines on Supplementary Figure 1.a and b. The Bragg reflection associated to this lattice containing a single π phase defect is represented in red on Supplementary Figure 1.c. The intensity drops down to zero exactly at the expected Bragg position. This is due to the destructive interference that occurs between the amplitudes scattered by the two out-of-phase volumes. Let us consider that the lattice has a length L, and that $A_1(q)$ is the amplitude scattered by the volume located between x = 0 and x = L/2, and $A_2(q)$ the amplitude scattered by the volume located between x = L/2 and x = L. The total scattered amplitude is then:

$$A(q) = A_1(q) + A_2(q).$$

$A_2(q)$ can be written in terms of $A_1(q)$ as:

$$A_2(q) = A_1(q)e^{inqa}\, e^{iqa/2} = A_1(q)e^{iqa/2},$$

where a is the lattice parameter of the perfect lattice. The expected amplitude at Bragg position q=2π/a is then:

$$A(2\pi/a) = A_1(2\pi/a) + A_1(2\pi/a)e^{i\pi} = A_1(2\pi/a) - A_1(2\pi/a) = 0.$$

For q values taken around 2π/a, the intensity is not 0, but follows the distribution displayed on the red curve of Supplementary Figure 1.c. Two secondary maxima are found on both sides of q=2π/a, with an intensity that is half of the Bragg intensity associated to the perfect lattice. Moreover, the width of the two red peaks is the same as that of the Bragg reflection associated to the perfect lattice. These features are characteristic to the presence of a π phase defect.

Dislocations are also π phase defects, but that develop in a three-dimensional lattice. Moreover, the phase variation can be smoother than the phase shift described in the case of the one-dimensional model. If one considers a perfect lattice

$$R(x,y,z) = \cos(2\pi x/a) + \cos(2\pi y/a) + \cos(2\pi z/a),$$

the associated Bragg reflection will display the same features as the one presented for a perfect one-dimension lattice. Let us introduce a dislocation in that lattice, for example an edge dislocation with a line along z, and a Burgers vector along x. The distorted lattice can then be described as:

$$R(x,y,z) = \cos(2\pi x/a+\varphi(x,y)) + \cos(2\pi y/a) + \cos(2\pi z/a),$$

where $\varphi(x,y)$ is the phase variation function accounting for the deformations induced by the dislocation. In the case of a dislocation line along z, with a Burgers vector along x, that function can be written as:

$$\varphi(x,y) = -\pi/2 * \mathrm{sgn}(y-y_0) + \tan^{-1}((K_y/K_x)^{1/2}*(x-x_0)/(y-y_0)),$$

where $(x_0,y_0)$ are the coordinates of the dislocation line position, and $K_x$ and $K_y$ are the lattice force constants of the system along x and y respectively. This formula describes well the displacement field around the dislocation line, but not the atomic positions in the dislocation core itself. The lattice containing such a dislocation in the middle of the volume, with $K_x = K_y = 1$, and the associated phase variations are represented Supplementary Figure 1.d and e. In that case, the phase variation is invariant around the dislocation line. Any cut of the phase distribution going through the dislocation line gives a similar phase profile as the one shown in red on Supplementary Figure 1.b. The calculated Bragg reflection thus reveals the same features as in the one-dimensional case, namely a zero intensity value at Bragg position, and a torus of maxima around it, showing that the destructive interference takes place in every direction of space. The torus has an intensity that is half the intensity of the Bragg reflection associated to the perfect lattice, and the same width. The torus appears perpendicular to the dislocation line direction, and any cut of this intensity distribution by a 2D detector will reveal a split reflection, unless the cut is exactly perpendicular to the z* direction.

The particular case of dissociated dislocation lines is discussed in the text. The principle is exactly the same, as the dissociated lines behave like a non-dissociated line in terms of phase around the lines. Only the stacking fault in-between the dissociated lines produces an additional signal.

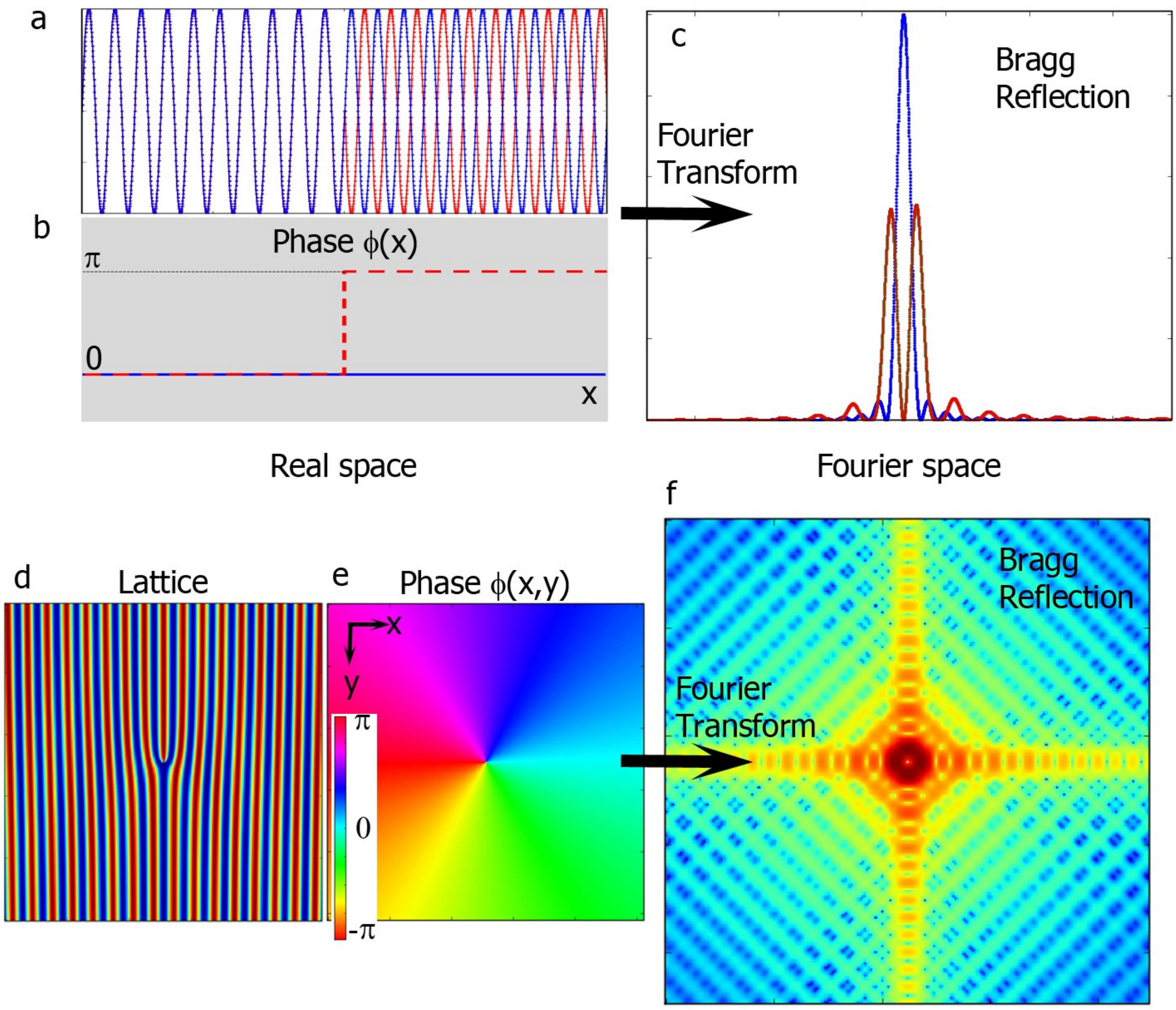

**Supplementary Figure 1 | Coherent diffraction pattern obtained when a π phase shift is present in real space. a**, One-dimensional periodic lattice with no phase defect (blue line), and with a π phase shift in the middle (red line). **b**, phase associated to the lattices shown on **a**. **c**, Bragg reflection associated to the perfect (blue line) and phase shifted (red line) lattices shown on **a**. When a π phase shift is present, a destructive interference leads to a intensity equal to 0 at the expected Bragg reflection position. The two maxima have same intensity, which is half the intensity of the Bragg reflection associated to the perfect lattice, and also the same width as it. **d**, Representation of a dislocation developping in a 2 dimensional elastically isotropic media, and **e**, associated phase. **f**, Intensity distribution around a Bragg reflection expected when the dislocation shown on **d** is present. Again, the intensity at Bragg position is exactly 0 due to the destructive interference, and a torus of intensity appears when the media is elastically isotropic.

## Supplementary Discussion 2

## Details about the intensity distribution seen on the 2D detector when the sample is slightly detuned from Bragg angle

In this experiment, the illumination of the defining 7 µm x 7 µm aperture inserted 8 cm before the sample produces diffracted beams, impinging the sample with a slightly different angle than the direct beam, although the sample is still in the near-field region. The calculated beam received by the sample is represented on Supplementary Figure 2 in linear and logarithmic scales perpendicularly to one set of blades.

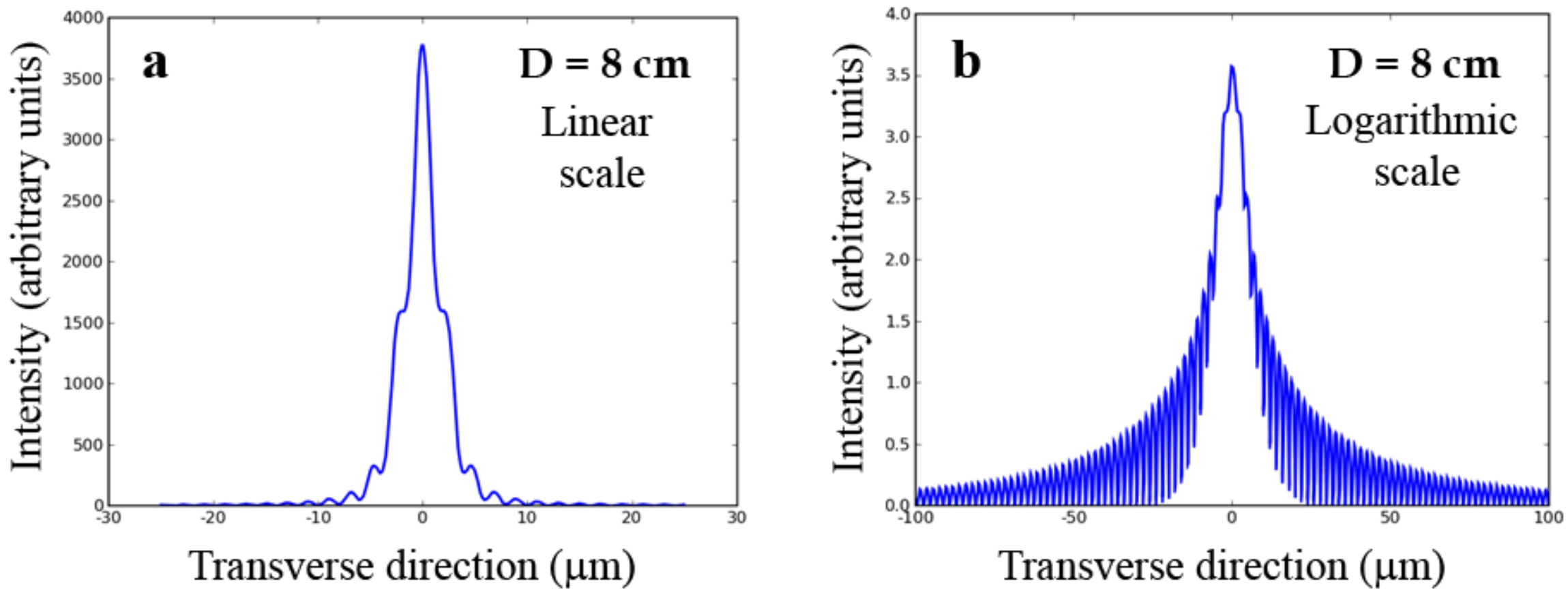

**Supplementary Figure 2 | Calculated beam profile at sample position.** At a distance D = 8 cm from the 7 µm x 7 µm slits, the sample is positioned at the limit between the near-field and far-field regions. Due to the diffraction by the slits, the beam profile perpendicular to the slits blades already displays some fringes, which are supplementary incident beams impinging the sample, and are able to be diffracted if they enter the Darwin width of the reflection. **a**, Linear scale. **b**, Logarithmic scale. Fringes up to 100 µm from the main beam are still visible.

The Ewald construction is represented on Supplementary Figure 3 for an incident angle equal to the (220) Bragg angle $\theta_{220}$. Supplementary Figure 3 shows that only the first order fringes appear in the vertical direction, but strongly attenuated. Indeed, the first order diffracted beam impinges the sample with an angle $\theta_{220}$ + 1.7 mdeg, which is still close to the (220) Bragg reflection condition (given the (111) monochromatized incident beam). The second order diffracted beam has an incident angle $\theta_{220}$ + 3.5 mdeg, which no longer satisfies the Bragg condition. That is why it does not appear in that direction on the CCD detector. On the contrary, the fringes that are perpendicular to the diffraction plane are diffracted up to a high order, due to the reduced angular selectivity of the sample in that direction.

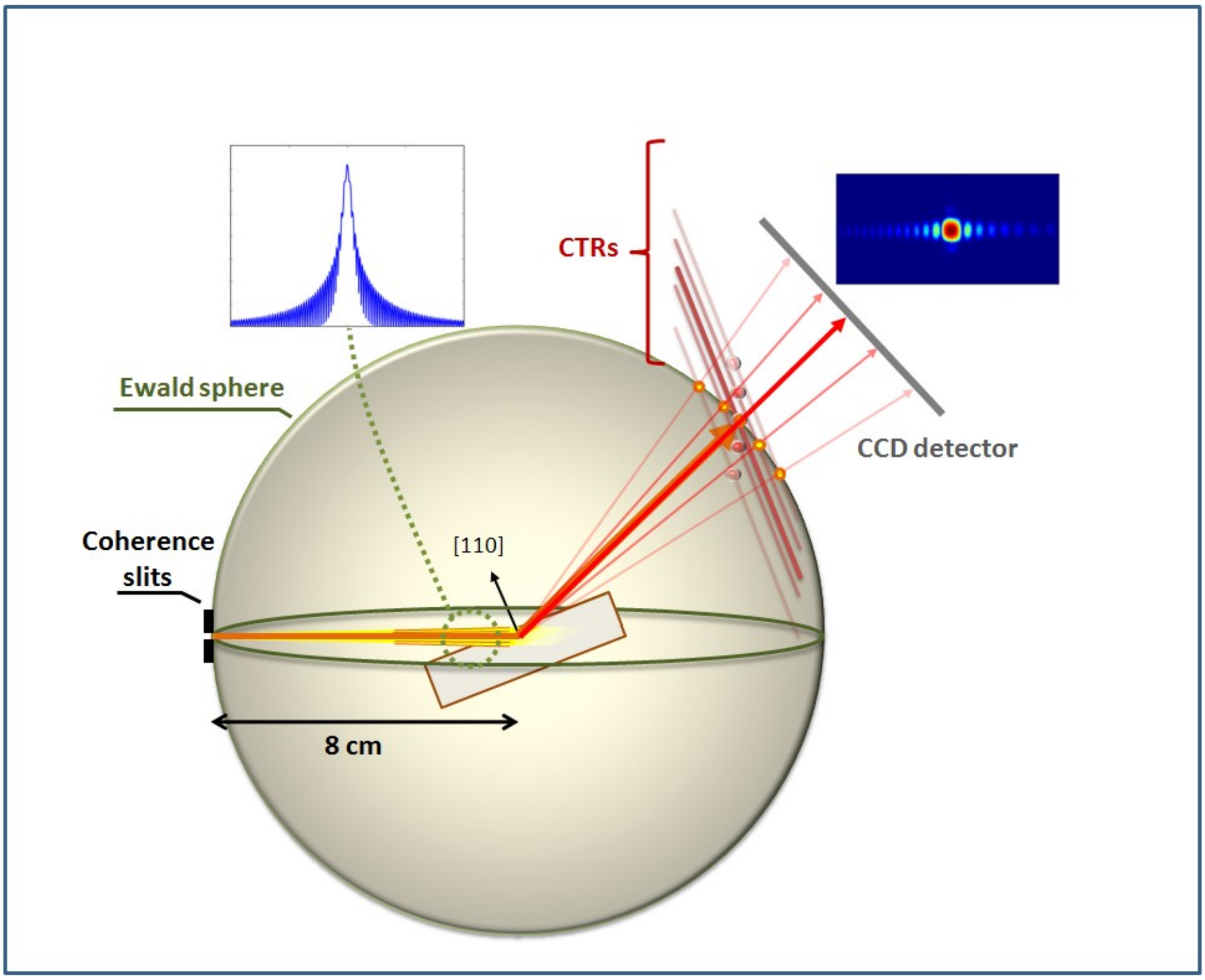

**Supplementary Figure 3 | Image seen on the 2D detector when the sample is in the Bragg geometry.** The sample is oriented in order to have the (220) reciprocal space node on the Ewald sphere. The supplementary beams due to the diffraction by the square slits defining the beam just before the sample are represented. These beams can be diffracted if they enter the Darwin width of the (220) reflection. The crystal truncation rod (CTR) decorates each spot in reciprocal space. The sketch only describes the behaviour in the diffraction plane, but the fringes and the associated CTRs are also present perpendicularly to the diffraction plane. Each spot or CTR that cut the Ewald sphere produces a signal on the 2D detector. A typical image obtained on the (220) Bragg reflection is shown. Only low order fringes appear in the diffraction plane, due to the small Darwin width of the (220) reflection. On the contrary, fringes appear up to high orders perpendicular to the diffraction plane, due to the lower selectivity in that direction.

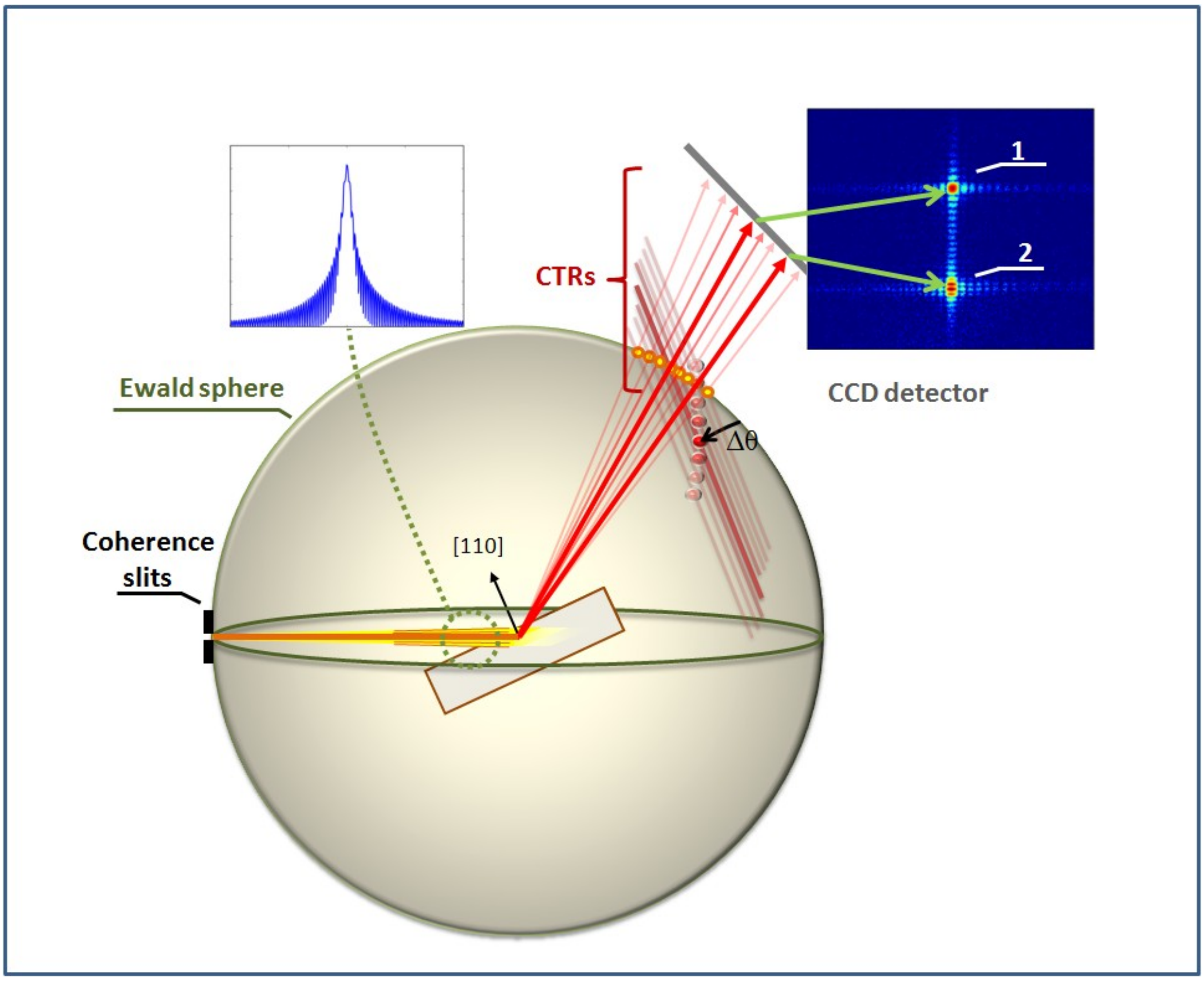

**Supplementary Figure 4 | Image seen on the 2D detector when the sample is slightly detuned from Bragg angle.** The (220) reciprocal space node is no longer cutting the Ewald sphere. It only appears through the associated CTR that produces a high signal on the 2D detector (1). Some diffracted beams coming from the diffraction by the slits satisfy the Bragg condition and produce fringes with a higher counting level (2). In between, the CTRs associated to the other diffracted beams coming from the diffraction by the slits appear periodically between (1) and (2). Perpendicular to the diffraction plane, the CTRs associated to the other diffracted beams coming from the diffraction by the slits also appear.